\documentclass[twocolumn,prl]{revtex4-1}
\usepackage{epsfig}

\begin{document}
\title{The nucleon and the two solar mass neutron star }
\author{Vikram Soni}
\affiliation{Centre for Theoretical Physics, Jamia Millia Islamia, New Delhi, India}
\author{Mitja Rosina}
\affiliation{Faculty of Mathematics and Physics, University of Ljubljana,
Jadranska 19, P.O. Box 2964, 1001 Ljubljana, Slovenia\\
J. Stefan Institute,  1000 Ljubljana, Slovenia}

\begin{abstract}
The existence of  a star  with such a large mass means that the equation
of state is stiff enough to provide a high enough pressure up to a fairly
large central densities,. Such a stiff equation of state is  possible if
the ground state has nucleons  as its constituents. This further implies
that a purely nucleon ground state may exist till  about  four times
nuclear density which indicates that  quarks in the nucleon are strongly
bound and that the nucleon nucleon potential is strongly repulsive.
We find this to be so in a chiral soliton model for the nucleon which
has bound state quarks. We point out that this has important implications
for the strong interaction  $ \mu_B$ vs T phase diagram.
\end{abstract}

\maketitle

\section{Introduction}
\label{introduction}

Equations of state (EOS) for stable stars like white dwarfs that involve
nonrelativistic electrons counteract gravitational infall of matter
through a fermi pressure that is proportional to the density to the
(5/3) power. When the mass of the star increases so does the electron density making
the electron fermi energy relativistic. Fermi pressures of relativistic
electrons  are proportional to density to the  (4/3) power and cannot
hold up to gravitational pressure. This is the Chandrasekhar instability
that sets the maximum mass of such  stars.

However, for neutron stars, even a pure nonrelativistic fermi gas of
neutrons is not sufficient to give large masses . Such a non interacting
nonrelativistic fermi gas can give stable neutron stars of maximum
mass about 0.7 solar mass -  this a general relativistic effect coming
from the Oppenheimer -- Volkoff equation. Beyond this mass  the 
pressure needs to be more repulsive than just the fermi pressure .  
This enhanced pressure is provided by nuclear interactions like
the hard core.

 It is  known that there are many purely nonrelativistic nucleon
 based neutron star models that have neutron stars with maximum mass
 above 2 solar masses, eg. the APR 98 EOS of Akmal, Pandharipande and
 Ravenhall \cite{APR}. For a brief review of nuclear stars and their
 EOS we refer the reader to \cite {Lattimer,mag}. It is also known
 that conventional
 hybrid stars with soft, relativistic quark matter cores surrounded by a
 nonrelativistic n+p+e plasma in beta equilibrium generally give a maximum mass
 for neutron stars of only $\sim$ 1.6 solar mass \cite{Lattimer,Soni1}.
 
The recent \cite{shapiro10} discovery of $\simeq 2 $ solar mass (highest
mass)  neutron star,
the recycled binary pulsar PSR J1614-2230,  using the precision technique
of Shapiro delay confronts us with question of what is the constituent
profile of such a star. This issue was first raised in \cite{SHR}

 In view of the foregoing, we investigate the following question; if
 matter in neutron stars is entirely composed
 of non relativistic nucleon degrees of freedom then can we have a simple
 resolution of this question?

It is useful to recall that the recycled binary pulsar, PSR J 1614-2230,
is rotating fast at a period of 3 millisec and we expect  a  $\sim$ 10\%
diminution of the central density from the rotation from centrifugal forces \cite{Haensel}. Since
APR 98 \cite{APR}   reports results for static stars,  we expect
the central density of a fast rotating 1.97 solar mass star to be
approximately $\sim$ the central density of a static 1.8 solar mass star.

\section{The Maxwell construction between nuclear matter and quark matter}

We work in an effective chiral symmetric theory that is QCD coupled to a
chiral sigma model. The theory thus preserves the symmetries of QCD. In
this effective theory chiral symmetry is spontaneously broken and the
degrees of freedom are constituent quarks which couple to colour singlet,
sigma and pion fields as well as gluons. Furthermore, since we  do not have 
exact solutions for a theory of the strong  interactions, we work in Mean field theory.
The �nucleon�  in such a theory
is a colour singlet quark soliton with three valence quark bound states
\cite{NPA}. The quark meson couplings are set by matching mass of the
nucleon to its experimental value and the meson self coupling is set from pi-pi scattering, which in turn sets
the tree level sigma particle mass  to be
of order 800 MeV. Such an effective theory has a range of validity up
to centre of mass energies ( or quark chemical potentials) of  $\sim$
800 MeV. For details we  refer the reader to ref. \cite{Soni1}.

This is one of the simplest effective  chiral symmetric theory for the strong
interactions at intermediate scale and we use this consistently to
describe, both, the composite nucleon of quark bound states and quark
matter. We expect it to be valid till the intermediate scales quoted
above. Of course inclusion of the higher mesonic degrees of freedom like
the  $\rho $  and  A1 would make for a more complete description. We work
at the mean field level where the gluon interactions are subsumed in the
colour singlet sigma and pion fields they generate. We could further
add perturbative gluon mediated corrections but they do not make an
appreciable difference.

One of lowest energy ground states at high baryon density that we find
in such chiral models is a neutral pion
 condensed state \cite{dautry,kutschera+90}. The equation of state for
 neutron stars  for such a state has been obtained in \cite{Soni1, Soni2}
 A simple way to look at whether nucleons can dissolve into quark matter
 is to plot  $E_B$,  the energy per baryon in the ground state of both,
 the quark matter
and the nuclear phases, versus $ 1/n_B$,  where $n_B$ is the baryon
density.  For the
 quark matter equation of state see Fig.1 \cite{Soni1} in which the
 quark matter EOS
 is indicated by the solid curves and the APR \cite{APR}  non relativistic
 nucleon
 EOS by the dashed line. The  slope  of the common tangent between the
 two phases then gives the pressure at the phase transition and the
 intercept, the common baryon  chemical potential.

\begin{figure}[htb]
\centering
\epsfig{file=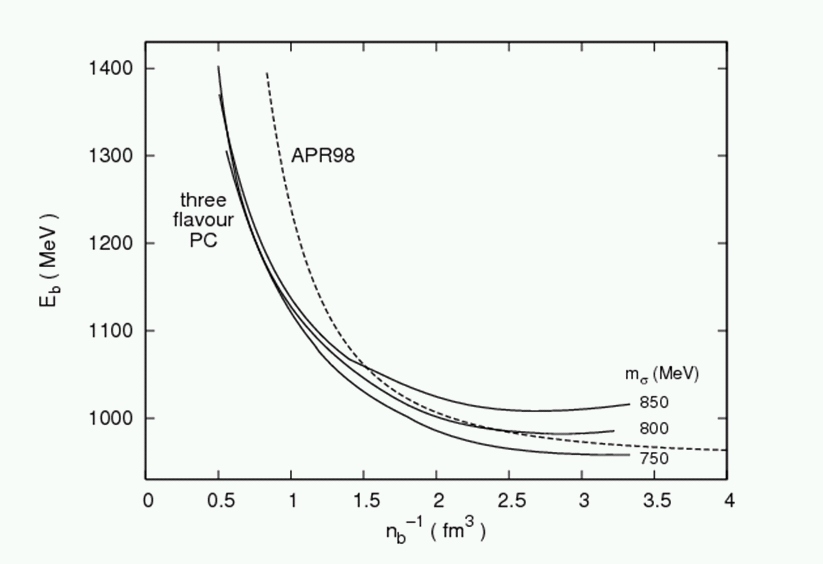,width=200pt}
\caption{The Maxwell construction: Energy per baryon plotted
against the reciprocal of the baryon number density for
APR98 equation of state (dashed line) and the 3-flavour pion-condensed
phase (PC) for three different values of  $m_\sigma$ (solid
lines). A common tangent between the PC phase and the
APR98 phase in this diagram gives the phase transition between
them. The slope of a tangent gives the negative of the
pressure at that point, and its intercept gives the chemical potential.
As this figure indicates, the transition pressure moves
up with increasing $m_\sigma$, and at $m_\sigma$ below $\sim$750 MeV a
common tangent between these two phases cannot be obtained.\\
(From Fig. 2 of Soni and Bhattacharya \cite{Soni1} ) }
\label{Fig1}
\end{figure}

As can be seen from Fig.1, it is the tree level value of the sigma mass
that determines the intersection of the two phases; the higher the mass
the higher the density at which the transition to quark matter will take
place. In \cite{Soni1}
 it was found that above, $m_\sigma\sim$ 850 MeV, stars with quark matter
 cores become unstable as their mass goes up  beyond the allowed maximum
 mass. So, if we want purely nuclear stars we should, in this model,
 work  at,  $m_\sigma \ge$  850 MeV  \cite{Soni1}.

 From Fig. 1, for the tree level value of the sigma mass $\sim$850 MeV,
 the common tangent in the two phases starts at  $ 1/n_B \sim  1.75$
 fm$^3$  ( $ n_B \sim 0,57$/fm$^3 $)   in the nuclear phase of
 APR   [A18 + dv +UIX] \cite{APR} and ends up at  $1/n_B \sim $  1.25
 fm$^3$  ($n_B \sim $ 0.8/ fm$^3$) in the quark matter phase.

At the above densities between the two phases there is a mixed phase
at the pressure given by the slope of the common tangent and the at a
baryon chemical potential given by the intercept of the common tangent
on the vertical axis. If we are to stay in the nuclear phase the best
way is to look at the  central density of the nuclear (APR) stars  and
if it so happens that the central density is lower  than that at which
the above phase transition begins the we can safely say that the star
remains in the nuclear phase.

Going back to the APR phase in in fig 11 of APR \cite{APR} we find that
for the APR  [A18 + dv +UIX] the central density of a star of 1.8 solar
mass is  $ n_B \sim$  0.62 /fm$^3$, very close to the initial density
at which the phase transition begins.

The reason we are taking a static star mass of 1.8 solar mass  from APR
\cite{APR} is that for PSR-1614, the star is rotating fast at a period
of 3 millisec and we expect  a  $\sim$ 10\% diminution of the central
density from the rotation \cite{Haensel}. Equivalently, since the above
paper reports results for static stars, the central density of a fast
rotating 1.97 solar mass star $\sim$ the central density of a static
1.8 solar mass star.

Now we have found that in above scenario the central density is of the
same order  as the density at which the above phase transition begins
in the nuclear phase.
Ideally we would like the central density to be a little less than
the initial density at which the above phase transition begins in the
nuclear phase.

\section{Beyond the Maxwell tangent construction for the phase transition}

How do we change the crossover and Maxwell tangent construction for the
phase transition?
There are two ways of moving the crossover between the 2 phases (and
also the initial density at which the above phase transition begins )
in the nuclear phase to higher density.

(i) By increasing the tree level mass of the sigma we can move the
quark matter curve up (Fig. 1),  thus moving the initial density at
which the above phase transition begins in the nuclear phase to higher
density. However we have to be careful. There is not much freedom here,
as this is what  also determines the $\pi-\pi$ scattering.

(ii)	At 3 - 4 times nuclear density $ n_B \sim$  0.6 /fm$^3$ the average energy per baryon 
in the APR EOS is less than 1050 MeV. However, the fermi energy  may be close to 
or above the lambda particle mass, which will make it possible to have
a small admixture of hyperons.

This will soften the nuclear EOS at high density. The phase transition
then begins
in the nuclear phase at higher density, but this will
also reduce the maximum mass. We have to ensure that the the maximum mass 
stays above two solar masses. There is an extesive literature on the effect of hyperons
at  well above  ( 3 -4 times) nuclear density but with no definite
conclusions. We thus do not pursue this question further here.


However, the Maxwell construction is not the final word on the phase
transition. In any case the above analysis assumes point particle nucleons. 
It does not take account of the structure and  the quark binding inside
the nucleon
( which depends mainly on the quark meson coupling ) or the nucleon
nucleon repulsion as we squeeze them. This is not captured by the Maxwell
construction. We now go on to show that this could move the transition from the 
nuclear to the quark phase to appreciably higher density.

\subsection{Binding energy of a quark in a nucleon}

An appoximate and simple expession for the energy of a colour singlet
nucleon soliton with three coloured bound state quarks is given below,
\cite{NPA}. Here , g is quark meson (Yukawa) coupling, $f_\pi$,
the pion decay constant and,  N, the number of bound state quarks.
We shall work with the dimensionless parameter, $X  =  R g f_{\pi}$,
where R is the soliton
radius. This follows from a simple parametrization for  a soluble model (see fig. 2). The
'mass' of a 'free' quark in this model is given  by,  $ m_q =   g f_{\pi}$,,

\begin{equation}
E/(g f_{\pi})  =  (\frac{3.12}{X} N  -  0.94.  N) + 24 \frac{X}{g^2}
\end{equation}

Minimizing this with respect to , X 

\begin{equation}
X^2   = \frac{3.12 g^2 N}{24}\\
\end{equation}

On substitution of this value
\begin{equation}
E_{min}/(g f_{\pi})  =  (\sqrt{\frac{3.12 N .24}{g^2} }) -  0.94N 
\end{equation}

For the nucleon we must set , $N = 3$ as all three quarks sit in the bound state.
We can now  evaluate the coupling,
g, by setting the nucleon mass to $ ~960 MeV $. This yields a value for ,
$ g \sim 6.9$.

However, the above formula allows us to look at  the energy of the configuration in 
which two quarks sit in the bound state  and one is moved up to the continum. Such a state
will give a measure of the energy required to unbind the nucleon. 

We note  that in this mean field model there is
no confinement: in any case, close to the quark matter phase transition
density, the nucleons will go into quark matter at the transition and
not 'free' quarks making confinement a peripheral issue.

We can easily check the possible bound states by evaluating the ratio
of  the  energy of  bound states with 2 and 3 quarks, which is given by,
$E_{min}/(g f_{\pi})$ and the respective number of  unbound  ( 'free' )quarks. The
'mass' of a 'free' quark in this model is given  by,  $ m_q =   g f_{\pi}$,
This is simply done by dividing the above equations by, N: if the answer
is less than ,1, we have a bound state, otherwise not.

\begin{eqnarray}
E_{min/}(N g f_{\pi})  &= & (\sqrt{\frac{3.12 N .24}{g^2} }) -  0.94N \nonumber\\
                       &\sim&  0.5 ~\text{for}~  N = 3\nonumber\\
		       &\sim& 0.83 ~\text{for}~  N =2\nonumber\\
	        	&\sim&  1.27~\text{for}~ N =1
\end{eqnarray}

indicating that regardless of the value of  $f_\pi $ we have bound states
for N = 2 and 3. Given the value of ,  $ g ~\sim 6.9$,  we can  find the
energy required to unbind a quark from such a nucleon. The energy of
a two quark bound state and an unbound quark is $1707 $ MeV  in comparision
to the energy of a 3 quark bound state nucleon which is , $~960 $ MeV.

i) The difference gives the binding energy of the quark in the nucleon,
$ ~  745 $ MeV.
The quark binding in this model is very high. In this model the quark
bound state eigenvalue (Fig. 2) \cite{NPA}  is well described by the
figure given below.

\begin{figure}[htb]
\centering
\epsfig{file=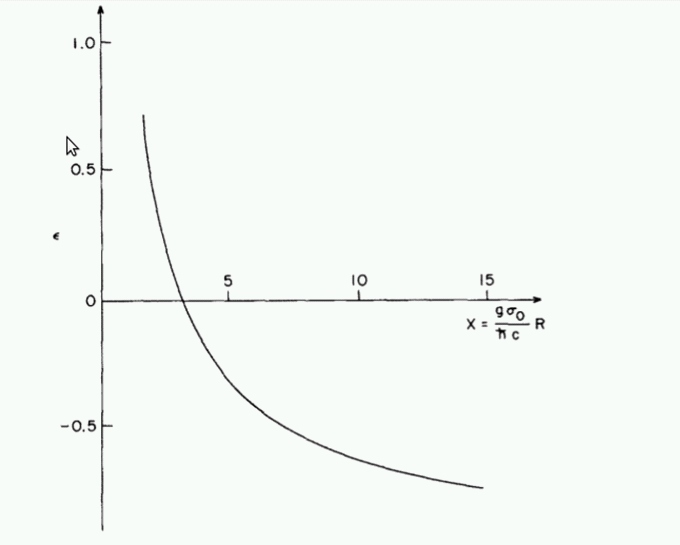,width=200pt}
\caption{Dependence of the quark energy on the soliton size $X$ in the
quark soliton model\\
 (From Fig. 2 of  Kahana, Ripka and Soni \cite{NPA}) }
\label{Fig2}
\end{figure}

ii) We can see that the quarks will become unbound ( go to the continuum)
when the energy eigenvalue is larger than the  unbound mass of the quark
which is given by  $m_q = g f_\pi$. This happens when  in
the dimensionless units used in Fig. 2

\begin{equation}
\epsilon\ge 1 ,   \text{at  X = 3.12/1.94 = 1.6}. 
\end{equation}
This translates into  $R$ =(1.6/2.5) fm$^{-1}\sim$ 0.6 fm$^{-1}$ .

This is the effective radius of the squeezed nucleon at which the
bound state quarks are liberated to the continuum. By inverting the
volume occupied by the nucleon and assuming hexagonal close packing,
this translates to nucleon density of

\begin{equation}
1/(6 R^3)\sim  0.77 fm^{-3}
\end{equation}

Thus the quark bound states in nucleon persist untill a much higher
density $\sim$ 0.8/fm$^3$. In other words, nucleons can survive well
above the density at which the Maxwell phase transition begins and
appreciably above the central density of the APR   2-solar-mass star.

\subsection{ Nucleon Nucleon repulsion}
Another feature is the  the nucleon nucleon potential. It has been found
for skyrmions and such quark solitons with skyrmion configurations that
there is a strong   N-N repulsion that forces the lowest baryon number
$N_\mathrm{B}= 2$ configuration to become toroidal \cite{Sriram}.
This is an indication that nucleon nucleon potential becomes strongly
repulsive.

It thus follows that the phase transition from nuclear to quark matter
will encounter a potential barrier before the quarks can go free.
This effect cannot be seen by the coarse Maxwell construction which does
not track their transition.

\section{ Discussion and Consequences }

These considerations will modify the simple minded Maxwell construction
above. It follows from above that the internal structure of the nucleon
will move
the phase transition to higher density.
All in all this produces a very plausible scenario of how the $\sim$2
solar mass star can be achieved in a purely nuclear phase.

A possible consequence of this unexpected scenario at high density is
that the the phase diagram of  QCD which plots temperature on the y axis
versus  baryon chemical potential on the
      x axis,  the quark matter transition for finite density ( in the
      range above) will be lifted up along the  temperature axis.

  For example, if for zero baryon density we have chiral restorstion at
$T_\Xi  \sim 150$ Mev, then at small baryon density such a temperature
will probably not be able to dissociate the nucleon bound state that
lives in a chirally broken (SBCS) ground state, as its binding energy
is  very large.   

Now we present a heuristic way of detrmining the  energy cost
of maintaining a nucleon as a quark bound state soliton (in an island
of spontaneously broken chiral symmetry) at this temperature. First, we
 estimate the thermal energy in a volume of a  nucleon of radius ,
$R  \sim 1 $ fermi, which is approximately,
$ \sim  \text{Volume} \cdot  {(kT_\Xi)}^4  \sim 250 $ Mev.
We must also add the cost in gradient energy
of decreasing
the meson VEV's  from  $ f_\pi $  inside of a soliton nucleon to  0
(the chiral symmetry restored value), outside of the nucleon. If we
assume this happens over a typical length scale of , $ \sim 1 $ fermi,
the gradient energy  also  works out to be,  $ \sim 200 - 250 $  Mev. The
sum of these energies is around $ 400- 500$ Mev, whereas ,the binding
energy of
the quark in such a nucleon is  $ \sim 750 $ Mev, indicating that  at
chiral restoration, $T_\Xi  \sim 150$ Mev, the nucleon may yet be intact.

 Thus, at finite but small baryon density  and $ T_\Xi \sim 150$ Mev,
 there may emerge a new intermediate mixed phase  in which nucleons will
 exist as bound states of locally spontaneously broken chiral symmetry
 (SBCS) in a sea of chirally restored quark matter.  Such a low baryon density 
state could be
seen in lattice calculations. This is quite the
 opposite to the popular bag notions of the nucleon  as being islands
 of restored chiral symmetry in a SBCS sea.

We also note that  in  the
chirally restored state,  the quarks acquire a typical temperature dependent
pole mass proprtional to, $ g T $  in perturbation theory where g is the QCD coupling constant.  The QCD coupling is still strong,
$ \frac{g^2 }{ 4\pi} \ge 1 $. For example, if at zero baryon density
we have chiral restoration at, $ T_\Xi \sim 150$ Mev, the 3 quarks that
make up a baryon have an , $ E_B \ge 1600$ Mev. This will also influence the 
transition to quark matter.

Acknowledgement: We are happy to acknowldge discussions and a partial collaboration with Pawel Haensel.

\end{document}